\documentclass[aps,prl, 
twocolumn, 
superscriptaddress,groupedaddress
,amsmath,amssymb
,floatfix
,nofootinbib
]{revtex4} 

\usepackage{graphicx}
\usepackage{dcolumn}
\usepackage{bm}
\usepackage[usenames,dvipsnames]{color}
\usepackage{amsmath,amssymb}
\usepackage{slashed}
\usepackage{hyperref}
\usepackage{float}
\usepackage[caption = false]{subfig}

\begin{document}

%

\let\a=\alpha      \let\b=\beta       \let\c=\chi        \let\d=\delta
\let\e=\varepsilon \let\f=\varphi     \let\g=\gamma      \let\h=\eta
\let\k=\kappa      \let\l=\lambda     \let\m=\mu
\let\o=\omega      \let\r=\varrho     \let\s=\sigma
\let\t=\tau        \let\th=\vartheta  \let\y=\upsilon    \let\x=\xi
\let\z=\zeta       \let\io=\iota      \let\vp=\varpi     \let\ro=\rho
\let\ph=\phi       \let\ep=\epsilon   \let\te=\theta
\let\n=\nu
\let\D=\Delta   \let\F=\Phi    \let\G=\Gamma  \let\L=\Lambda
\let\O=\Omega   \let\P=\Pi     \let\Ps=\Psi   \let\Si=\Sigma
\let\Th=\Theta  \let\X=\Xi     \let\Y=\Upsilon
%


\def\cA{{\cal A}}                \def\cB{{\cal B}}
\def\cC{{\cal C}}                \def\cD{{\cal D}}
\def\cE{{\cal E}}                \def\cF{{\cal F}}
\def\cG{{\cal G}}                \def\cH{{\cal H}}
\def\cI{{\cal I}}                \def\cJ{{\cal J}}
\def\cK{{\cal K}}                \def\cL{{\cal L}}
\def\cM{{\cal M}}                \def\cN{{\cal N}}
\def\cO{{\cal O}}                \def\cP{{\cal P}}
\def\cQ{{\cal Q}}                \def\cR{{\cal R}}
\def\cS{{\cal S}}                \def\cT{{\cal T}}
\def\cU{{\cal U}}                \def\cV{{\cal V}}
\def\cW{{\cal W}}                \def\cX{{\cal X}}
\def\cY{{\cal Y}}                \def\cZ{{\cal Z}}

%

\newcommand{\Ns}{N\hspace{-4.7mm}\not\hspace{2.7mm}}
\newcommand{\qs}{q\hspace{-3.7mm}\not\hspace{3.4mm}}
\newcommand{\ps}{p\hspace{-3.3mm}\not\hspace{1.2mm}}
\newcommand{\ks}{k\hspace{-3.3mm}\not\hspace{1.2mm}}
\newcommand{\des}{\partial\hspace{-4.mm}\not\hspace{2.5mm}}
\newcommand{\desco}{D\hspace{-4mm}\not\hspace{2mm}}


\def\be{\begin{equation}}
\def\ee{\end{equation}}
\def\bea{\begin{eqnarray}}
\def\eea{\end{eqnarray}}
\def\bm{\begin{matrix}}
\def\em{\end{matrix}}
\def\bpm{\begin{pmatrix}}
	\def\epm{\end{pmatrix}}

{\newcommand{\lsim}{\mbox{\raisebox{-.6ex}{~$\stackrel{<}{\sim}$~}}}
{\newcommand{\gsim}{\mbox{\raisebox{-.6ex}{~$\stackrel{>}{\sim}$~}}}
\def\mpl{M_{\rm {Pl}}}
\def\gev{{\rm \,Ge\kern-0.125em V}}
\def\tev{{\rm \,Te\kern-0.125em V}}
\def\mev{{\rm \,Me\kern-0.125em V}}
\def\ev{\,{\rm eV}}

\title{\boldmath Neutrino dark energy and leptogenesis with TeV scale triplets}
\author{Chandan Hati}
\email{chandan@prl.res.in}
\affiliation{ Physical Research Laboratory, Navrangpura, Ahmedabad 380 009, India.}
\affiliation{Indian Institute of Technology Gandhinagar, Chandkheda, Ahmedabad 382 424, India.}
\author{Utpal Sarkar}
\email{utpal@prl.res.in}
\affiliation{ Physical Research Laboratory, Navrangpura, Ahmedabad 380 009, India.}
\date{\today}
\begin{abstract}
We propose a realization of mass varying neutrino dark energy in two extensions of the Standard Model (SM) with a dynamical neutrino mass related to the acceleron field while satisfying the naturalness. In the first scenario the SM is extended to include a TeV scale scalar Higgs triplet ($\xi$) and a TeV scale second Higgs doublet ($\eta$), while in the second scenario an extension of the SM with fermion triplet $(\Sigma)$ is considered. We also point out the possible leptogenesis mechanisms for simultaneously generating the observed baryon asymmetry of the universe in both the scenarios and discuss the collider signatures for the $\tev$ scale new fields which make these models testable in the current run of LHC. 

\end{abstract}

\maketitle

Evidences from the astrophysical observations suggest that out of the total mass-energy budget of the universe, the baryonic and dark matter together account for only about 30\% while the remaining 70\%, referred to as dark energy, is attributed to the accelerated expansion of the universe and remains a challenge to explain. While the existence of a scalar field called quintessence provides an explanation, a striking proximity of the effective scales of neutrino masses and the dark energy points to a connection between them, realized in the neutrino dark energy ($\nu$DE) models. To this end, several approaches have been used in the literature. In some scenarios, a direct connection through the formation of neutrino condensate at a late epoch of the early universe using the effective self-interaction has been explored \cite{Kapusta:2004gi}. While another class of models invoke the variation of neutrino masses to dynamically obtain the dark energy \cite{Gu:2003er, Fardon:2003eh, Hung:2000yg, Ma:2006mr, Afshordi:2005ym}. In this Letter, we will focus on this latter approach. 

The atmospheric, solar and reactor neutrino oscillation experiments have confirmed the existence of tiny but nonzero masses of neutrinos ($\sim 10^{-2} \ev$). An attractive explanation of neutrino masses employs the seesaw mechanism \cite{Minkowski:1977sc}, giving rise to naturally small Majorana or Dirac masses of neutrinos. In addition, the baryon asymmetry of the universe can be generated through leptogenesis \cite{Fukugita:1986hr} in the framework of the seesaw scenario. In the original $\nu$DE models, the Standard Model (SM) is extended to accommodate singlet right-handed neutrinos ($N_i$, $i=1,2,3$) giving a Majorana mass to light neutrinos. The Majorana mass of the right-handed neutrinos are made to vary with the acceleron field, connecting the light neutrino masses with the scale of dark energy. However, the naturalness requires the Majorana masses of the right-handed neutrinos to be in the $\ev$ range, in contradiction to the expected scenario of a very heavy $M_{N_i}$ triggering the canonical seesaw mechanism. In Ref. \cite{Ma:2006mr}, it was pointed out that the above problem can be avoided if the SM is extended to include triplet Higgs scalars. However, in such a scheme the coefficient of the trilinear scalar coupling with mass dimension varies with the acceleron field and this predicts the mass scale of the triplet Higgs scalars to be close to the electroweak symmetry breaking scale (of order $100 \gev$), which has not been observed at the LHC so far. 

The purpose of this Letter is to point out two ways to get around the above constraint, while simultaneously explaining the observed baryon asymmetry of the universe. One way is to add some additional scalar field to push the additional scalar field masses to $\tev$ scale, readily testable at the current run of LHC. Another way is to add fermion triplets instead of scalar triplets and utilize the type III seesaw scheme. 

First we propose a realization of mass varying neutrinos in an extension of the usual triplet Higgs model which includes a second Higgs doublet ($\eta$) in addition to the SM Higgs doublet ($\Phi$) and Higgs triplet ($\xi$), but no right-handed neutrinos \cite{Ma:1998dx, Ma:2002nn}. In this scenario both the additional Higgs fields are of $\tev$ scale and the smallness of neutrino mass comes from the lepton number breaking scalar sector. This model has highly predictive collider signatures and thus it can be right away put to test in the current run of LHC. Next we propose a new model of $\nu$DE utilizing an extension of the SM with fermion triplet $(\Sigma^{+}, \Sigma^{0}, \Sigma^{-})_{R}$, where the neutrino mass is dynamical and related to the acceleron field. This model can naturally give the correct energy scale associated with the neutrino mass and it provides a rich $\tev$ scale phenomenology, testable at the LHC. We also point out possible leptogenesis mechanisms for simultaneously generating the observed baryon asymmetry of the universe in both the models.

By extending the SM to include a heavy Higgs triplet ($\xi^{++}$, $\xi^{+}$, $\xi^{0}$) with trilinear couplings to both the lepton doublet $L_{i}=(\nu_{i}, l_{i})$ and the Higgs doublet $\Phi = (\phi^{+}, \phi^{0})$, one can realize the unique dimension-five effective operator \cite{Weinberg:1979sa}
\be{\label{1.1}}
\cL_{\rm{eff}}=\frac{f_{ij}}{\Lambda} L_{i} L_{j}\Phi \Phi,
\ee
obtained by integrating out the heavy degrees of freedom (with mass much larger than the ordinary SM particles) associated to a characteristic heavy mass scale $\Lambda$. Thus the neutrinos, massless in the minimal SM, acquire small Majorana masses. The relevant interaction terms are given by
\bea{\label{1.2}}
\cL_{\rm{int}} &=& f_{ij}\left[ \nu_{i}\nu_{j}\xi^{0} +\frac{1}{\sqrt{2}} (\nu_{i} l_{i} +l_{j} \nu_{j})\xi^{+} +l_{i} l_{j}\xi^{++}\right]\nonumber\\
&& + \rm{h.c.}\; ,
\eea
where $\tilde{\Phi}=(\bar{\phi}^{0}, -\phi^{-})$. The above interaction terms give \cite{Ma:1998dx}
\be{\label{1.3}}
\left(M_{\nu}\right)_{ij}=\frac{2 f_{ij} \mu \langle \phi^{0}\rangle ^{2}}{m_{\xi^{0}}^{2}}.
\ee
Thus it follows that if $\mu$ is a function of the acceleron field $\cA$ i.e. $\mu=\mu (\cA)$, then the mass varying neutrinos can be realized for $m_{\xi}$ of the order of the electroweak scale. However, if the $\nu$DE is indeed realized through the Higgs triplet, then at least $\xi^{++}$ should have been observable at the LHC. Thus it is worth exploring if such a Higgs triplet can be schemed to have a mass of $\tev$ scale in light of the current run of LHC.

{\bf{Model A:}} In presence of the additional Higgs doublet $\eta$ in the  above scheme, the neutrino masses come from the Higgs triplet $\xi$ (with lepton number assignment $L=-2$) and its interaction with $\eta$ (carrying lepton number $L=-1$) \cite{Ma:2002nn}. The most general lepton number conserving scalar potential is given by
\bea{\label{1.4}}
V &=& m_{1}^{2} \Phi^{\dagger}\Phi +m_{2}^{2}\eta^{\dagger}\eta +m_{3}^{2} \rm{Tr} [\Delta^{\dagger} \Delta] \nonumber\\
    &+& \frac{1}{2}\lambda_{1} (\Phi^{\dagger}\Phi)^{2} + \frac{1}{2}\lambda_{2} (\eta^{\dagger}\eta)^{2} + \frac{1}{2}\lambda_{3} (\rm{Tr} [\Delta^{\dagger} \Delta])^{2}\nonumber\\
    &+&\frac{1}{2}\lambda_{4} (\rm{Tr} [\Delta^{\dagger} \Delta^{\dagger}]) (\rm{Tr} [\Delta \Delta]) + \lambda_{5} (\Phi^{\dagger}\Phi)(\eta^{\dagger}\eta)\nonumber\\
    &+&\lambda_{6} (\Phi^{\dagger}\Phi)(\rm{Tr} [\Delta^{\dagger} \Delta])+\lambda_{7} (\eta^{\dagger}\eta) (\rm{Tr} [\Delta^{\dagger} \Delta])\nonumber\\
    &+& \lambda_{8} (\Phi^{\dagger}\eta)(\eta^{\dagger}\Phi)+\lambda_{9}(\Phi^{\dagger} \Delta^{\dagger}\Delta\Phi)+\lambda_{10}(\eta^{\dagger}\Delta^{\dagger}\Delta\eta)\nonumber\\
    &+& \mu(\eta^{\dagger}\Delta \tilde{\eta}) +\rm{h.c.} \; ,
\eea 
where 
\be{\label{1.5}}
\Delta=\bpm \xi^{+} / \sqrt{2} & \xi^{++} \\ \xi^{0} & - \xi^{+} / \sqrt{2}\epm ,
\ee  
 $\tilde{\eta}=(\bar{\eta}^{0}, -\eta^{-})$ and $\mu$ has the dimension of mass. The lepton number is softly broken by the terms
 \be{\label{1.6}}
 V_{\rm{soft}}=\mu_{1}^{2} \Phi^{\dagger}\eta+\mu_{2} \left(\Phi^{\dagger}\Delta\tilde{\eta}\right)+\mu_{3}\left(\Phi^{\dagger}\Delta\tilde{\Phi}\right)+\rm{h.c.} \; .
 \ee
Next we define vacuum expectation values (VEVs) of the scalar fields to be $\langle\phi^{0}\rangle=v_{1}$,  $\langle\eta^{0}\rangle=v_{2}$ and $\langle\xi^{0}\rangle=v_{3}$. Now minimization of the potential with respect to the various Higgs fields give the consistency conditions and the relations between the different VEVs, which can be solved assuming $m_{1}^{2}<0$, but $m_{2}^{2}>0$ and $m_{3}^{2}>0$ to obtain
\bea{\label{1.7}}
v_{1}^{2} &\simeq& -m_{1}^{2}/\lambda_{1},\nonumber\\
v_{2} &\simeq& -\mu_{1}^{2}v_{1}/[m_{2}^{2}+(\lambda_{5}+\lambda_{8})v_{1}^{2}],\nonumber\\
v_{3} &\simeq& - \left(\mu v_{2}^{2}+\mu_{2}v_{1}v_{2}+\mu_{3} v_{1}^{2}\right)/(m_{3}^{2}+\lambda_{6}v_{1}^{2}).
\eea
Thus taking $m_{2}$, $m_{3}$ and $\mu$ to be $M \sim \tev$ we have
\be{\label{1.8}}
v_{2}\sim \mu_{1}^{2}v_{1}/M^{2}, \; \; \; \; \; v_{3}\sim v_{2}^{2}/M. 
\ee
Consequently, it follows that $u\ll v_{2}\ll v_{1}$ and 
\be{\label{1.9}}
v_{3} \sim \mu_{1}^{2} v_{1}^{2}/M^{5}.
\ee
For $v_{1}\sim 10^{2} \gev$ and $\mu_{1}\sim 1 \gev$ we have $v_{2}\sim 0.1 \mev$ and $v_{3}\sim 10^{-2}\ev$, which gives the correct order of magnitude for neutrino mass $(m_{\nu})_{ij}=2 f_{ij} v_{3}$. Thus we have a natural realization of the required small neutrino masses with $\tev$ scale additional Higgs fields, which does not need any large extra space dimensions constraining $m_{\xi}$ below the cutoff energy scale. Moreover this model is much more flexible compared to the scenario with only Higgs triplet in the sense that there is no strict constraint on $m_{\xi}$ to be of the order of electroweak scale. Now the realization of $\nu$DE model trough mass varying neutrinos is straight forward. The idea is to make $\mu_{1}$ a function of the acceleron field $\cA$, i.e. $\mu_{1}=\mu_{1}(\cA)$. We will come back to the realization of $\nu$DE once we give the account of the other model below.

{\bf{Model B:}} The extension of the fermion (lepton) sector of the SM can be realized in two ways. The basic idea is that the new lepton multiplet gains a large mass and then it mixes with the ordinary lepton doublet triggering the seesaw mechanism. The new lepton multiplet can only be a singlet or a triplet of $SU(2)_{L}$. The idea of the triplet lepton representation to utilize the seesaw structure in neutrino mass matrix was first proposed in Ref. \cite{Foot:1988aq}, referred to as type III seesaw, which have been generalized in the context of unified theories in Ref. \cite{Barr:2003nn}. The simplest way to utilize type III seesaw is to add the $SU(2)_{L}$ triplet with zero hypercharge
\be{\label{2.1}}
\Sigma=\bpm \Sigma^{0}/\sqrt{2} & \Sigma^{+}\\ \Sigma^{-} & -\Sigma^{0}/\sqrt{2}\epm,
\ee
 to the SM, with the interaction terms
 \be{\label{2.2}}
 \cL_{\rm{int}}=-\frac{1}{2} \rm{Tr}\left[ \bar{\Sigma} M_{\Sigma} \Sigma^{c}+ \bar{\Sigma}^{c} M_{\Sigma}^{\ast}\Sigma\right] -\tilde{\Phi}^{\dagger} \bar{\Sigma}\sqrt{2} Y_{\Sigma} L -\bar{L}\sqrt{2}Y^{\dagger}_{\Sigma}\Sigma \tilde{\Phi}.
 \ee
The terms related to neutrino mass matrix can be identified readily to obtain the mass matrix as
\be{\label{2.3}}
\cL_{\nu, \rm{mass}}=\bpm \nu & \Sigma^{0} \epm \bpm 0 & Y_{\Sigma} v/2\sqrt{2} \\ Y_{\Sigma}^{T} v/2\sqrt{2} & M_{\Sigma}/2 \epm \bpm \nu \\ \Sigma^{0} \epm. 
\ee
This gives the non zero neutrino masses given by
\be{\label{2.4}}
M_{\nu}=-\frac{v^{2}}{2}Y_{\Sigma}^{T}M_{\Sigma}^{-1}Y_{\Sigma}.
\ee
Now the neutrino masses can be connected to the dark energy by simply taking $M_{\Sigma}=M_{\Sigma}(\cA)$, however such a scenario is constrained from naturalness. This scheme can be generalized right away by accommodating the right handed neutrinos $N^{c}_{i}, i=1,2,3$ in the scenario. The most general neutrino mass matrix in such a scenario can be written as
\be{\label{2.5}}
\cL_{\rm{\nu, mass}}=\bpm \nu & N^{c} & \Sigma^{0} \epm 
\bpm 0 & M_{N} & F_{1} u \\ M_{N}^{T} &0 & F_{2}\Omega \\ F^{T}_{1}u & F^{T}_{2}\Omega & M_{\Sigma} \epm
\bpm \nu \\ N^{c} \\ \Sigma^{0} \epm, 
\ee
where the off-diagonal terms in the third column and row correspond to the mass terms ${F_{1}}_{ij} \nu_{i} \Sigma^{0} u$ and ${F_{2}}_{ij} N^{c}_{i} \Sigma^{0} \Omega$ with $u$ and $\Omega$ being the VEVs of the corresponding Higgs fields. The realization of nonrenormalizable term giving rise to type III seesaw, by integrating out heavy fields is shown in Fig. \ref{fig1}.
\begin{figure}[h]
	\includegraphics[width=8cm]{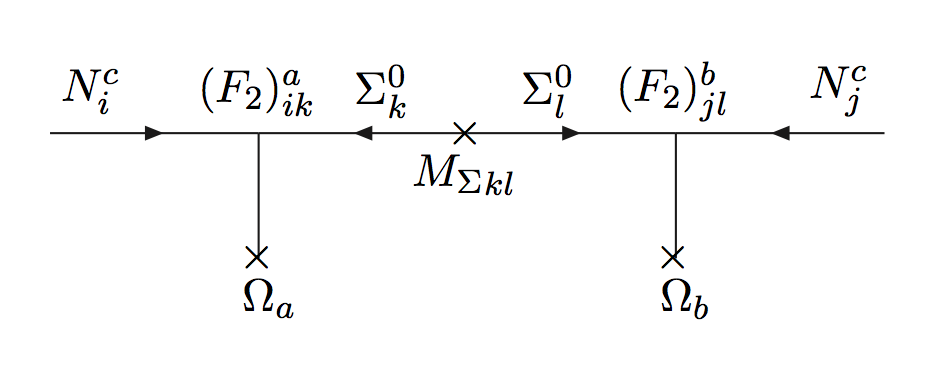}
	\caption{Diagram realizing the effective nonrenormalizable operator generating right handed neutrino mass $M_{R}=(F_{2}\Omega) M_{\Sigma}^{-1} (F_{2}^{T}\Omega)$.}
	\label{fig1}
\end{figure}
 The neutrino mass in the above scenario have two kind of contributions, given by
\be{\label{2.6}}
M_{\nu}=-M_{N} (F_{2}\Omega M_{\Sigma}^{-1} F_{2}^{T}\Omega)^{-1} M_{N}^{T}-(M_{N}+M_{N}^{T})\frac{u}{\Omega},
\ee
where the first term corresponds to a ``double seesaw'' contribution and the second term corresponds to the type III seesaw contribution. The relative contributions of the two kinds of terms to $M_{\nu}$ is model dependent. To satisfy the naturalness constraints we will consider the case $M_{N}\sim 1 \ev$. Now taking $u\sim v\sim 10^{2}\gev$, $\Omega \sim 10^{4} \gev$ and considering the phenomenologically interesting case $M_{\Sigma}\sim 10^{3} \gev$ with verifiable implications at the current run of LHC, it follows that for $F_{2}\gtrsim 10^{-6}$ the dominant contribution to $M_{\nu}$ in Eqn. (\ref{2.6}) comes from the second term associated with the type III seesaw contribution and for the above set of values we obtain $M_{\nu}\sim 10^{-2} \ev$ as desired. The mass varying neutrinos can be realized by taking $M_{N}=M_{N}(\cA)$.

Having given the details of the two models realizing mass varying neutrinos with desired small masses while satisfying the naturalness, we are now ready to discuss the realization of $\nu$DE where the neutrino mass (assumed to be a function of the canonically normalized acceleron field $\cA$) $M_{\nu}(\cA)$ is a dynamical quantity and $\partial M_{\nu}/\partial \cA \neq 0$ \cite{Fardon:2003eh}. In the nonrelativistic limit, the energy density consists of the thermal neutrino (and antineutrino) background ($M_{\nu} n_{\nu}$) and the scalar potential $V_{0}(M_{\nu})$. The effective potential can be written as
\be{\label{3.1}}
V(M_{\nu})=M_{\nu}n_{\nu}+V_{0}(M_{\nu}).
\ee
The neutrino background (driving $M_{\nu}$ to small values) gets diluted as the universe expands and the source term decreases as a result, while $V_{0}$ is minimized for a large $M_{\nu}$. Thus the two terms act in the opposite directions with a minimum at some intermediate $M_{\nu}$ with a nonzero $V_{0}$. The minimum of the effective potential is given by
\be{\label{3.2}}
V^{\prime}(M_{\nu})=n_{\nu}+V^{\prime}_{0}(M_{\nu})=0.
\ee
Now at any instant of time assuming the simple equation of state
\be{\label{3.3}}
p(t)=\omega \rho(t),
\ee
it follows that
\be{\label{3.4}}
\omega +1 =-\frac{\partial \log V}{3\partial \log a}=-\frac{M_{\nu}V_{0}^{\prime}(M_{\nu})}{V}=\frac{\Omega_{\nu}}{\Omega_{\nu}+\Omega_{\cA}},
\ee
where $\Omega_{\nu}=M_{\nu}n_{\nu}/\rho_{c}$ is the neutrino energy density and $\Omega_{\cA}=\rho_{A}/\rho_{c}$ corresponds to the contribution of $V_{0}(M_{\nu})$ to the energy density, with $\rho_{c}$ is the critical density and $a$ is the cosmic scale factor. Since the observed value of $\omega\simeq -1$, Eqn. (\ref{3.4}) implies that the energy density in the thermal neutrino background must be much less compared to the total dark energy density. This in turn suggests that the potential $V_{0}(M_{\nu})$ should be a flat potential. For the case where $d\omega/d n_{\nu}$ is small, the relation
\be{\label{3.5}}
M_{\nu}\propto n_{\nu}^{\omega}
\ee
holds. The above considerations are independent to any specific model of neutrino mass \cite{Fardon:2003eh} and we will use them to draw out the phenomenological consequences specific to the two models of interest.

As we have discussed above in model A, $\mu_{1}=\mu_{1}(\cA)$ makes the effective mass of the neutrinos to vary. While in model B, $M_{N}=M_{N}(\cA)$ does the same. Now for the self interactions of the acceleron field $\cA$ we take the effective potential of the form
\be{\label{3.6}}
V_{0}=\Lambda^{4} \log (1+|\cM(\cA)/\bar{\cM}|),
\ee
where in model A, $\cM(\cA)=\mu_{1}(\cA)$ and in model B, $\cM(\cA)=M_{N}(\cA)$. Hence Eqn. (\ref{3.1}) takes the form
\be{\label{3.7}}
V(x)=a_{1} x+ a_{2} \log \left(1+\frac{x}{a_{3}}\right),
\ee
where $x=M_{\nu}\propto |\mu (\cA)|$ and $a_{1}$, $a_{2}$, $a_{3}$ are all positive. Now assuming $\cM(\cA)/\bar{\cM}\gg 1$ it follows that $x_{\rm{min}}\propto a_{2}/a_{1}$ implying
\be{\label{3.8}}
M_{\nu}\propto n_{\nu}^{-1} ,
\ee
which gives the desired $\omega\simeq-1$. Thus, the two models under consideration can naturally explain the $\nu$DE for $\tev$ scale $\xi$, $\eta$ masses in model A and $\tev$ scale mass of the new fermion triplet $\Sigma$ in model B. The $\tev$ scale mass of these particles makes these model particularly interesting in the context of collider phenomenology at the LHC. We will come back to the implications and signatures of these two models for colliders such as the LHC, once we address the issue of leptogenesis in these two models. 

In model A, the SM is extended to include scalar triplet and an additional Higgs doublet $\eta$, providing an attractive possibility of realizing a sucessful leptogenesis scenario. We start with the conventional formalism of scalar triplet leptogenesis in a hierarchical case. $SU(2)_{L}\times U(1)_{Y}$ is the valid gauge group at an energy scale far above the electroweak symmetry breaking. Thus it follows that if we analyze one of the three components of the triplet scalar field then the results will hold for the other two. From Eqs. (\ref{1.2}), (\ref{1.4}) and (\ref{1.6}) we can read off the decay modes of $\xi^{++}$ as
\be{\label{4.1}}
\xi^{++}_{a} \rightarrow  \left\{ \bm l^{+}_{i} l^{+}_{j} & (L=-2), \\ \phi^{+} \phi^{+} & (L=-2), \\ \eta^{+} \eta^{+} & (L=-0). \em \right. 
\ee
The coexistence of the above decay modes implies nonconservation of lepton number, however, the lepton asymmetry generated by $\xi^{++}$ gets compensated by the decays of $\xi^{--}$, unless $CP$ is also violated and the decays take place out-of-equilibrium. We follow the mass matrix formalism \cite{Ma:1998dx, Flanz:1994yx}, where the tree level mass matrix for the triplets are assumed to be real and diagonal. Hence $CP$ is conserved at tree level, however $CP$ conservation occurs at one-loop level due to interference between the tree and one-loop diagrams shown in Fig. \ref{fig2}.
\begin{figure}[h]
	\includegraphics[width=8cm]{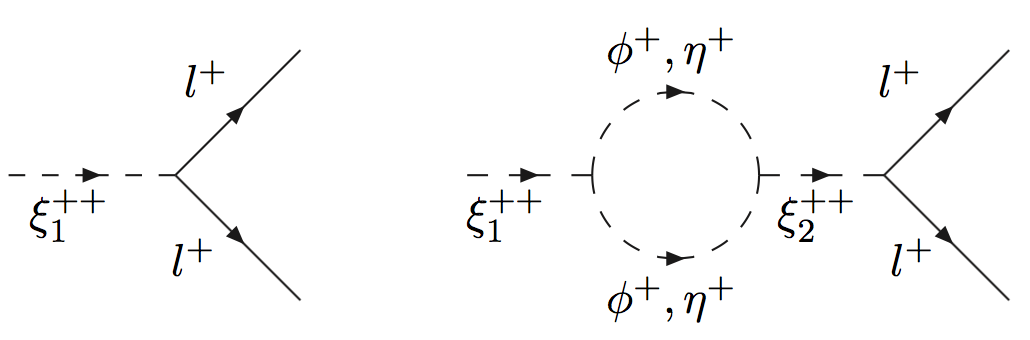}
	\caption{The tree level (left) and one-loop (right) decay diagrams for $\xi^{++}\rightarrow l^{+}l^{+}$. A lepton asymmetry is generated by the $CP$ violation occurring due to the interference between them.}
	\label{fig2}
\end{figure}
 Note that, at least two $\xi$'s are required for $CP$ nonconservation to occur. Following the mass-matrix formalism of Refs. \cite{Flanz:1994yx}, the diagonal tree-level mass matrix of $\xi_{a}$ in Eqn. (\ref{1.4}) is modified in the presence of interactions to
\be{\label{4.2}}
\frac{1}{2}\xi^{\dagger} \left(M_{+}^{2}\right)_{ab}\xi_{b}+ \frac{1}{2} \left(\xi^{\ast}_{a}\right)^{\dagger}\left(M_{-}^{2}\right)_{ab}\xi^{\ast}_{b},
\ee
where
\be{\label{4.3}}
M_{\pm}^{2}=\bpm M_{1}^{2}-i\Gamma_{11}M_{1} & -i \Gamma_{12}^{\pm}\\ -i \Gamma_{21}^{\pm}M_{1} &M_{2}^{2}-i\Gamma_{22}M_{2}\epm ,
\ee
with $\Gamma_{ab}^{+}=\Gamma_{ab}$ and $\Gamma_{ab}^{-}=\Gamma_{ab}^{\ast}$. From the absorptive part of the one loop diagram for $\xi_{a}\rightarrow \xi_{b}$ we have
\be{\label{4.4}}
\Gamma_{ab}M_{b}=\frac{1}{8\pi} \left( \mu_{a}\mu_{b}^{\ast}+{\mu_{3}}_{a}{\mu_{3}}_{b}^{\ast}+M_{a}M_{b}\sum_{k,l}f^{\ast}_{akl}f_{bkl}\right).
\ee
Now for $\Gamma_{a}\equiv \Gamma_{aa}\ll M_{a}$, the eigenvalues of $M_{\pm}^{2}$ are given by
\be{\label{4.5}}
\lambda_{1,2}=\frac{1}{2} (M_{1}^{2}+M_{2}^{2}\pm \sqrt{S}),
\ee
where $S=(M_{1}^{2}-M_{2}^{2})^{2}-4 |\Gamma_{12} M_{2}|^{2}$ and $M_{1}>M_{2}$. The physical states are given by
\be{\label{4.6}}
\psi^{+}_{1,2}=a_{1,2}^{+}\xi_{1}+b^{+}_{1,2}\xi_{2}\; , \hspace{0.3 in} \psi^{-}_{1,2}=a_{1,2}^{-}\xi_{1}^{\ast}+b^{-}_{1,2}\xi_{2}^{\ast}\; ,
\ee 
where $a_{1}^{\pm}=b_{2}^{\pm}=1/\sqrt{1+|C_{i}^{\pm}|^{2}}$, $b_{1}^{\pm}=C_{1}^{\pm}/\sqrt{1+|C_{i}^{\pm}|^{2}}$, $a_{2}^{\pm}=C_{2}^{\pm}/\sqrt{1+|C_{i}^{\pm}|^{2}}$ with
\bea{\label{4.7}}
C_{1}^{+}=-C_{2}^{-} &=& \frac{-2i\Gamma^{\ast}_{12}M_{2}}{M_{1}^{2}-M_{2}^{2}+ \sqrt{S}} \; ,\nonumber\\
C_{1}^{-}=-C_{2}^{+} &=& \frac{-2i\Gamma_{12}M_{2}}{M_{1}^{2}-M_{2}^{2}+ \sqrt{S}} \; .
\eea
The states $\psi^{\pm}_{1,2}$ evolve with time and decay into lepton pair and antilepton pair \footnote{Note that $\xi_{a}$ and $\xi_{a}^{\ast}$ are $CP$ conjugate states, while $\psi^{\pm}_{i}$ are not.}. Assuming $(M_{1}^{2}-M_{2}^{2})^{2} \gg 4|\Gamma_{12} M_{2}|^{2}$, the lepton asymmetries generated are given by \cite{Ma:1998dx}
\bea{\label{4.8}}
\e_{i}=\frac{1}{8\pi^{2}(M_{1}^{2}-M_{2}^{2})^{2}}\sum_{k,l}\left\{ {\rm{Im}} \left[ \mu_{1}\mu_{2}^{\ast} f_{1kl}f_{2kl}^{\ast}\right]\right.\nonumber\\
+ \left. {\rm{Im}} \left[ {(\mu_{3})}_{1}{(\mu_{3})}_{2}^{\ast} f_{1kl}f_{2kl}^{\ast}\right] \right\} \left[ \frac{M_{i}}{\Gamma_{i}}\right].
\eea
For the case $M_{1}>M_{2}$, when the temperature of the universe cools down below $M_{1}$, $\psi_{1}$ decays away to create a lepton asymmetry. However this asymmetry is washed out by lepton number nonconserving interactions of $\psi_{2}$ and the subsequent decay of $\psi_{2}$ at a temperature below $M_{2}$ sustains. The lepton asymmetry then gets converted to baryon asymmetry in the presence of the anomalous $B+L$ violating processes before the electroweak phase transition. The approximate final baryon asymmetry generated is given by
\be{\label{4.9}}
\frac{n_{B}}{s}\sim \frac{\e_{2}}{3 g^{\ast} K(\ln K)^{0.6}},
\ee
where $K\equiv \Gamma_{2}(M_{2}/T =1)/H(M_{2}/T =1)$ is a parameter measuring the deviation from thermal equilibrium, with the Hubble rate defined by $H=1.66 {g^{\ast}}^{1/2}(T^{2}/M_{\rm{Pl}})$, where $g^{\ast}$ corresponds to the number of relativistic degrees of freedom.

Other than the decays and the inverse decays of triplet scalars, one needs to incorporate the gauge scatterings $\psi\bar{\psi} \leftrightarrow F \bar{F}, \phi \bar{\phi}, G \bar{G}$ ($F$ corresponds to SM fermions and $G$ corresponds to gauge bosons) and $\Delta L=2$ scattering processes $ll\leftrightarrow \phi{\ast}\phi^{\ast}$ and $l \phi\leftrightarrow \bar{l}\phi^{\ast}$ into the Boltzmann equation analysis of the asymmetry. Including the above washout processes, it turns out that $M_{\xi}\gtrsim 10^{11}\gev$ is required in order to generate the correct asymmetry \cite{Hambye:2005tk}. However, for a quasi-degenerate spectrum of scalar triplets the resonance effect can enhance the CP-asymmetry by a large amount and a successful leptogenesis scenario can be attained for a much smaller value of triplet scalar mass. A detailed analysis of the resonant leptogenesis is beyond the scope of this Letter and an account of the same can be found in Ref. \cite{Strumia:2008cf}, where an absolute bound of $M_{\xi}\gtrsim 1.6 \tev$ is obtained for a successful leptogenesis scenario.

In model B, the type III seesaw scheme is realized and the right handed neutrino mass matrix enters in the formula for light neutrino masses compared to the type I seesaw. As a consequence, the light neutrino masses, mixing and leptogenesis are not that tightly coupled as in the case of type I seesaw, where the constraints on the right-handed neutrino mass $M_{R}$ can clash with the constraints coming from the textures of light neutrino masses and mixings. The advantage of the type III seesaw mechanism given in Eqn. (\ref{2.5}) is that, instead of the three heavy Majorana neutrinos in type I seesaw, here we have six heavy Majorana neutrinos. This can give rise to three pseudo-Dirac pairs of neutrinos with one or more pairs having degenerate masses. The six heavy two component neutrinos have the form of the mass matrix given by \cite{Albright:2003xb}
\be{\label{5.1}}
\bpm \tilde{N}^{c}_{i} & \tilde{\Sigma}^{0}_{i} \epm 
\bpm 0 & M_{i}\delta_{ij} \\ M_{i}\delta_{ij} & \tilde{M_\Sigma}_{ij}  \epm
\bpm \tilde{N}^{c}_{j} \\ \tilde{\Sigma}^{0}_{j} \epm. 
\ee
Now the degenerate lightest pair of pseudo-Dirac neutrinos or equivalently, two Majorana neutrinos $N_{\pm}\simeq (\tilde{N}^{c}_{1}\pm \tilde{\Sigma}^{0}_{1})/\sqrt{2}$ with masses $M_{\pm}\simeq \pm M_{1}+\frac{1}{2}\tilde{M_\Sigma}_{11}$ can decay into light neutrino and Higgs doublet via the Yukawa term $Y_{i\pm}(N_{\pm} \nu_{i})\Phi$, where
\be{\label{5.2}}
Y_{i\pm}\simeq \frac{(\tilde{Y}_{i1}\pm (\tilde{F}_{2})_{i1})}{\sqrt{2}}\mp \frac {\tilde{M}_{11}}{4M_{1}}\frac{(\tilde{Y}_{i1}\mp (\tilde{F}_{2})_{i1})}{\sqrt{2}}.
\ee
The asymmetry generated by the decays of $N_{\pm}$ is given by
\be{\label{5.2.1}}
\e_{1}=\frac{1}{4\pi}\frac{{\rm{Im}} \left[ \sum_{j}(Y_{j+} Y^{\ast}_{j-}) \right]^{2}}{\sum_{j} \left( |Y_{j+}|^{2}+|Y_{j-}|^{2} \right)} I(M_{-}^{2}/M_{+}^{2}),
\ee
where $I(M_{-}^{2}/M_{+}^{2})$ comes from the absorptive part of the decay amplitude, with $I(x)=\sqrt{x}\left[ 1-(1+x)\ln(1+(1/x))+1/(1-x) \right]$. Using the new basis parametrization  $N^{c}_{i}=U_{ij}\tilde{N}^{c}_{j}$ and ${\Sigma_{0}}_{i}=V_{ij} \tilde{\Sigma}_{0j}$ with the matrix $(F_{1})_{ij}$ diagonal, where
\be{\label{5.3}}
U=\bpm u_{11} &\lambda u_{12}&\lambda u_{13}\\ \lambda u_{21} & u_{22}& u_{23}\\ \lambda u_{31} & u_{32}& u_{33} \epm,
\ee
with $u_{ij}\sim 1$ and
\be{\label{5.4}}
\tilde{F}_{2} u=\bpm \lambda^{2}f_{11} &\lambda f_{12}&\lambda f_{13}\\ \lambda f_{21} & f_{22}& f_{23}\\ \lambda f_{31} & f_{32}& f_{33} \epm v_{u},
\ee
the asymmetry can be put in the form \cite{Albright:2003xb}
\be{\label{5.5}}
\e_{1}=\frac{\lambda^{2}}{4\pi}\frac{\left( |u_{31}|^{2}-|f_{31}|^{2} \right) {\rm{Im}}(u^{\ast}_{31} f_{31})}{|u_{31}|^{2}+|f_{31}|^{2}+|f_{21}|^{2}} I.
\ee
The lepton asymmetry of the universe is computed using
\be{\label{5.6}}
Y_{L}=\frac{n_{B}}{s}\sim \frac{\e_{2} d}{ 3 g^{\ast} K(\ln K)^{0.6}},
\ee
where $d$ is the washout parameter. In this case, for a hierarchical mass spectrum of triplets the lower bound on triplet mass for a successful leptogenesis scenario is given by $M_{\Sigma}\gtrsim 3\times 10^{10}$ \cite{Hambye:2003rt, Fischler:2008xm} and to have a $\tev$ scale leptogenesis one must assume a quasi-degenerate spectrum of fermion triplets giving resonant enhancement as in the case of scalar triplets, giving $\tev$ scale bound on $M_{\Sigma}$ \cite{Strumia:2008cf}.

The triplet fields $\xi$ and $\Sigma$ can be produced at the LHC if their masses are of the order of $\tev$ and therefore, LHC gives an unique opportunity to verify the mechanism of neutrino mass generation if any of these heavy states or their signatures are observed. To this end, we give a very brief summery of the production and observability of the triplet fields in the two models discussed above. A quantitative exploration of the discovery potential of these new fields is beyond the scope of this Letter and here we mainly concentrate on a qualitative account of the likely scenarios.

The members of the scalar triplet field can be produced at the LHC via the channels
\bea{\label{6.1}}
q\bar{q} &\rightarrow& Z^{\ast}/\gamma^{\ast}\rightarrow \xi^{++}\xi^{--},\nonumber\\
q_{1}\bar{q}_{2} &\rightarrow& {W^{\pm}}^{\ast} \rightarrow \xi^{++}\xi^{\mp},\nonumber\\
q\bar{q} &\rightarrow& Z^{\ast}/\gamma^{\ast}\rightarrow \xi^{+}\xi^{-}.
\eea
In the above three channels the interactions are fixed by the triplet gauge couplings and hence the production cross sections only depend on the scalar masses. In addition to the above three channels, there are additional channels where the scalar triplet field can be produced in association with $W^{\pm}$ or quarks
\bea{\label{6.2}}
q_{1}\bar{q}_{2} &\rightarrow& {W^{\pm}}^{\ast} \rightarrow \xi^{++}W^{\mp},\nonumber\\
q_1 q_2 &\rightarrow& {W^{\pm}}^{\ast} {W^{\pm}}^{\ast} q_3 q_4 \rightarrow \xi^{\pm\pm}q_3 q_4, \nonumber\\
q_1 q_2 &\rightarrow& Z^{\ast} Z^{\ast} q_3 q_4 \rightarrow \xi^{\pm\pm}q_3 q_4, \nonumber\\
q_1 q_2 &\rightarrow& \gamma^{\ast} \gamma^{\ast} q_3 q_4 \rightarrow \xi^{\pm\pm}q_3 q_4.
\eea
The associated production with $W^{\pm}$ and single production via $W^{\pm}W^{\pm}$ fusion involve the $\xi^{\pm\pm}W^{\pm}W^{\pm}$ vertex, which is suppressed by a factor $\e=v_{3}/v_{1}$. The $\gamma\gamma$ and $ZZ$ fusion processes are also very suppressed compared to the pair production cross section. 

The possible $\xi^{\pm \pm}$ decay modes are
\bea{\label{6.3}}
\xi^{\pm\pm} &\rightarrow& l^{\pm}_{i} l^{\pm}_{j}, \nonumber\\
\xi^{\pm\pm}&\rightarrow& W^{\pm} W^{\pm}, \nonumber\\
\xi^{\pm\pm}&\rightarrow& \xi^{\pm} W^{\pm}, \nonumber\\
\xi^{\pm\pm}&\rightarrow& \xi^{\pm} \xi^{\pm},
\eea
where $l_{i}=e, \mu, \tau$ for $i=1, 2, 3$. The decay mode into pair of leptons have been extensively discussed in the literature because it provides a clear multi-lepton final state signatures for the pair production of doubly charged Higgs field with a very small SM background \cite{delAguila:2008cj}. The possible two body decay modes of $\xi^{\pm}$ are
\bea{\label{6.4}}
\xi^{\pm}&\rightarrow& l^{\pm}_{i} \nu_{j}, \nonumber\\
\xi^{\pm}&\rightarrow& W^{\pm} Z, \nonumber\\
\xi^{\pm}&\rightarrow& u_{j} \bar{d}_{k}, \bar{u}_{j} d_{k},
\eea
with the last two decay modes again suppressed by a factor $\e=v_{3}/v_{1}$. Thus the production of scalar triplet fields can give rise to several possible final states. The final states can be classified according to the number of charged leptons as (a) $l^{+}l^{+}l^{-}l^{-}X$, (b) $l^{\pm}l^{\pm}l^{\mp}X$, (c) $l^{\pm}l^{\pm}X$, (d) $l^{+}l^{-}j_{\tau} X$, (e) $l^{\pm} j_{\tau} j_{\tau} j_{\tau} X$, where $l$ corresponds to electrons or muons (not necessarily all with the same flavor), $j_{\tau}$ corresponds to a tau jet and $X$ represents additional jets   \cite{delAguila:2008cj}. The unique signature of model A is the decay mode $\xi^{++}\rightarrow \eta^{+}\eta^{+}$, if kinematically allowed.

Similarly, in model B the dominant partonic production channels of the charged and neutral components of the fermion triplet are given by
\bea{\label{7.1}}
q\bar{q} &\rightarrow& Z^{\ast}/\gamma^{\ast}\rightarrow \Sigma^{+}\Sigma^{-},\nonumber\\
q_{1}\bar{q}_{2} &\rightarrow& {W^{\pm}}^{\ast} \rightarrow \Sigma^{\pm}\Sigma^{0}.
\eea
The decay modes of $\Sigma^{\pm}, \Sigma^{0}$ are
\bea{\label{7.2}}
\Sigma^{\pm}&\rightarrow& l^{\pm} Z, \nonumber\\
\Sigma^{\pm}&\rightarrow& l^{\pm} \Phi, \nonumber\\
\Sigma^{\pm}&\rightarrow& \bar{\nu} W^{+}, \nu W^{-}, \nonumber\\
\Sigma^{0}&\rightarrow& l^{\pm} W^{\mp}, \nonumber\\
\Sigma^{0}&\rightarrow& \nu Z, \nonumber\\
\Sigma^{0}&\rightarrow& \nu \Phi.
\eea
 Here the final states with different no of leptons can be classified as (a) six leptons, (b) five leptons, (c) $l^{\pm}l^{\pm}l^{\pm}l^{\mp}X$, (d) $l^{+}l^{+}l^{-}l^{-}X$, (e) $l^{\pm}l^{\pm}l^{\pm}X$, (f) $l^{\pm}l^{\pm}l^{\mp}X$, (g) $l^{+}l^{-}X$, (h) $l^{+}l^{-} j j j j X$ and (i) $l^{\pm} j j j j X$ \cite{delAguila:2008cj, Franceschini:2008pz}. The unique signatures of type III seesaw such as six lepton and five lepton final states can be used to distinguish it from the type II seesaw scheme at the LHC.

In conclusion, we have studied the realization of mass varying neutrinos in an extension of the usual triplet Higgs model by including an extra Higgs doublet ($\eta$) and an extension of the SM with fermion triplet $(\Sigma^{+}, \Sigma^{0}, \Sigma^{-})_{R}$. We find that both the scenarios can accomodate $\nu$DE with a dynamical neutrino mass related to the acceleron field while satisfying the naturalness, in the former scenario with $\tev$ scale triplet Higgs fields ($\xi$) and additional doublet Higgs field $(\eta)$ and in the latter scenario with $\tev$ scale fermion triplets $\Sigma$. We also point out the possible leptogenesis mechanisms for simultaneously generating the observed baryon asymmetry of the universe in both the scenarios. Finally, the $\tev$ scale new fields in both the models give unique and highly predictive collider signatures, testable in the current run of LHC. 
 
\end{document}